# Surface conduction of topological Dirac electrons in bulk insulating Bi$_2$Se$_3$

Dohun Kim*[1], Sungjae Cho*[1,†], Nicholas P. Butch[1,‡], Paul Syers[1], Kevin Kirshenbaum[1], Shaffique Adam[2], Johnpierre Paglione[1], and Michael S. Fuhrer[1,+]

1. *Center for Nanophysics and Advanced Materials, Department of Physics, University of Maryland, College Park, MD 20742-4111, USA*

2. *Center for Nanoscale Science and Technology, National Institute of Standards and Technology, Gaithersburg, MD 20899-6202, USA*

**The newly-discovered three-dimensional strong topological insulators (STIs) exhibit topologically-protected Dirac surface states[1,2]. While the STI surface state has been studied spectroscopically by e.g. photoemission[3-5] and scanned probes[6-10], transport experiments[11-17] have failed to demonstrate the most fundamental signature of the STI: ambipolar metallic electronic transport in the topological surface of an insulating bulk. Here we show that the surfaces of thin (<10 nm), low-doped Bi$_2$Se$_3$ (≈10$^{17}$/cm$^3$) crystals are strongly electrostatically coupled, and a gate electrode can completely remove bulk charge carriers and bring both surfaces through the Dirac point simultaneously. We observe clear surface band conduction with linear Hall resistivity and well-defined ambipolar field effect, as well as a charge-inhomogeneous minimum conductivity region[18-20]. A theory of charge disorder in a Dirac band[19-21] explains well both the magnitude and the variation with disorder strength of the minimum conductivity (2 to 5 $e^2/h$ per surface) and the residual (puddle) carrier density (0.4 x 10$^{12}$ cm$^{-2}$ to 4 x 10$^{12}$ cm$^{-2}$). From the measured carrier mobilities 320 cm$^2$/Vs to 1,500 cm$^2$/Vs, the charged impurity densities 0.5 x 10$^{13}$ cm$^{-2}$ to 2.3**

**x $10^{13}$ cm$^{-2}$ are inferred. They are of a similar magnitude to the measured doping levels at zero gate voltage (1 x $10^{13}$ cm$^{-2}$ to 3 x $10^{13}$ cm$^{-2}$), identifying dopants as the charged impurities.**

*these authors contributed equally to this work.

Bi$_2$Se$_3$ as prepared is observed to be *n*-type due to Se vacancies. We find that mechanically exfoliated thin (thickness $t \approx$ 10 nm) Bi$_2$Se$_3$ on SiO$_2$/Si is invariably highly *n*-doped with sheet charge densities $\gtrsim 10^{13}$ cm$^{-2}$, much greater than expected considering the bulk charge density ($\approx 10^{17}$ cm$^{-3}$) in our low-doped starting material[22], suggesting additional doping is induced by mechanical cleavage, reaction with ambient species[14,23], or substrate interaction. In order to remove this doping, we employed two types of *p*-type doping schemes on mechanically-exfoliated thin Bi$_2$Se$_3$ field-effect transistors on 300 nm SiO$_2$/Si back gate substrates[24-27]: (1) chemical doping with 2,3,5,6-tetrafluoro-7,7,8,8-tetracyanoquinodimethane (F4TCNQ) or (2) electrochemical doping with a polymer electrolyte top gate.

Figure 1a and b show schematics of device structures and gating schemes. We exploit either strong electron affinity ($\approx$ 5.4 eV) of F4TCNQ molecules[27] or large capacitance ($\approx$ 1μF/cm$^2$) of the electrochemical double layer at the interface between accumulated ions and the sample surface[24-26] to induce negatively charged ions near the surface and *p*-type doping of Bi$_2$Se$_3$. In both cases the dopant density was fixed after cooling to cryogenic temperature, but further tuning of the carrier density was possible using the back gate (see Methods).

Figure 1c and d show the longitudinal resistivity $\rho_{xx}$ and Hall carrier density $n_H = 1/(eR_H)$ (where $R_H$ is the Hall coefficient, and $e$ is the elementary charge) of a representative device (F4TCNQ-doped device 4) at various temperatures $T$ from 2K to 50K as a function of back gate voltage $V_g$. $\rho_{xx}(V_g)$ shows a peak at approximately $V_{g,0}$ = -45 V, and $n_H$ changes sign at a similar $V_g$, diverging positively (negatively) when approaching $V_{g,0}$ = -45 V from above (below). There is no evidence of an energy gap: $\rho_{xx}(T)$ is metallic ($d\rho_{xx}/dT > 0$) and saturates at low $T$, and $n_H(T)$ shows little temperature dependence. The behavior is strongly reminiscent of that seen for the two-dimensional Dirac electronic system in graphene[18]. Likewise, we identify the linear regions of $n_H$ vs $V_g$ for $V_g$ > -35 V and $V_g$ < -60 V as unipolar *n*- and *p*-doped regimes respectively, and the region -35 < $V_g$ < -60 V as an inhomogeneous regime where electron and hole transport are both present.

Figure 2a shows the Hall resistivity as a function of magnetic field $\rho_{xy}(B)$ of device 4 at various gate voltages in the unipolar *n*- and *p*-doped regimes. The Hall resistivity in the unipolar regime is always linear over the entire range of magnetic field (±9 T), indicating all bands contributing to the transport have similar mobility and same carrier sign. Specifically, we can rule out the possibility of both bulk and surface channels participating in conduction (previously observed to give a non-linear $\rho_{xy}(B)$ [11, 12, 28]) or significant contribution to conduction by impurity bands which should have much lower carrier mobilities (1 cm$^2$/Vs to 10 cm$^2$/Vs) [17]. The measured linear ambipolar Hall effect with carrier mobility of >10$^3$ cm$^2$/Vs is therefore a strong indication that conduction in our samples is dominated by the surface states.

We note that a previous work on singly-gated $Bi_2Se_3$ of similar thickness but heavily (0.5%) Ca-doped[17] also showed a superficially similar resistivity peak, interpreted there as the transition from bulk to surface conduction. No region of unipolar *p*-type Hall effect was observed. The authors concluded that significant band-bending in these highly-doped crystals led to very different carrier densities on either side of the device as well as an effective reduction of the bulk gap[11,13,17]. In order to determine whether band bending is important in our devices, we fabricated an additional top gate on an F4TCNQ-doped device (device 5), using hydrogen silsequioxane (HSQ) as a top-gate dielectric.

Figure 2b shows the resistivity $\rho_{xx}$ of dual-gated device 5 as a function of applied displacement field to the top ($D_{tg}$) and bottom ($D_{bg}$) surfaces (see Methods). The data are presented as a polar plot of normalized resistivity ($\rho_{xx}/\rho_{max}$) of the device as a function of total magnitude of displacement field $D_{total}=|D_{bg} + D_{tg}|$ and asymmetry factor defined by $\alpha = (4/\pi)\tan^{-1}[(D_{tg} - D_{bg})/D_{total}]$. We find that the measured resistivity depends only on *total* displacement field, proportional to the total charge density in the $Bi_2Se_3$ slab. We conclude from the observed azimuthal symmetry that both surfaces are gated simultaneously with either gate, and their chemical potential lies at the same level. If the gates acted independently on top and bottom surfaces, the maximum resistivity peak associated with the transition from *n*- to *p*-doping in each surface should broaden or split with increasing asymmetry; no such effect is observed. Remarkably, we find that simultaneous gating can be achieved even with a single gate electrode ($\alpha = \pm1$). We ascribe this effect to (1) the strong electrostatic coupling of the surfaces due to the large intersurface capacitance provided by the thin, lightly doped $Bi_2Se_3$, which has a high relative dielectric constant $\kappa \approx 100$; and (2) the small density of states of the Dirac surface. The net result is that the electrostatic intersurface capacitance exceeds the quantum capacitance of each surface, in which case the two surface potentials become locked together (see Supplementary Information for more discussion).

Having eliminated the possibilities of band-bending or significant contribution to the conductivity by bulk or impurity states, we conclude that our measurement probes the conductance of the simultaneously-gated ambipolar Dirac surfaces states. Our results therefore represent the first experimental demonstration of metallic, ambipolar, gapless electronic conduction of the topological surface state in $Bi_2Se_3$ in the absence of bulk carriers, the defining quality of a topological insulator.

Below, we analyze in more detail the transport properties of the topological surface state as a function of carrier density per surface $n$ estimated from $n = (C_g/2e)(V_g - V_{g,0})$ where $V_{g,0}$ is the gate voltage at which $R_H = 0$, which corresponds closely to the gate voltage of minimum conductivity. Figure 3 shows (a) conductivity per layer ($\sigma$), (b) the Hall carrier density per layer ($n_{H,layer} = 1/2R_H e$), and (c) field effect mobility $\mu = \sigma/ne$ vs. carrier density per layer $n$. Data are shown for devices 1-3 with electrolyte gating, and devices 4-5 charge transfer doped with F4TCNQ. Several features are notable immediately in Figure 3 and comprise the major experimental observations in this work. Upon carrier density tuning, (1) $\sigma$ and $n_{H,layer}$ show clear ambipolar conduction with well-defined $p$- and $n$- regions, (2) $|n_{H,layer}|$ shows a minimum value ($n^*$) for $p$- and $n$- conduction, (3) $\sigma$ shows a roughly linear carrier density dependence for $n^* < n < n_{bulk}$ where $n_{bulk} \approx 5 \times 10^{12}/cm^2$ is the carrier density above which the bulk conduction band is expected to be populated, and (4) a minimum conductivity ($\sigma_{min} = 2\ e^2/h$ to $5\ e^2/h$) is observed.

Extending the theory of charge disorder in graphene[19,20], a recent theoretical study predicts the conductivity limited by charged-impurity scattering in STI of the form (assuming linear Dirac band)[21],

$$\sigma(n) \sim C \left|\frac{n}{n_{imp}}\right| [e^2/h] \quad \text{for } n > n^* \qquad (1a)$$

$$\sigma(n) \sim C \left|\frac{n^*}{n_{imp}}\right| [e^2/h] \quad \text{for } n < n^* \qquad (1b)$$

where $n_{imp}$ is the charged impurity density, $C$ is a constant which depends on the Wigner-Seitz radius $r_s$, and $n^*$ is identified as the residual carrier density in electron and hole puddles. For Bi$_2$Se$_3$ on SiO$_2$[21] we expect $0.05 < r_s < 0.2$ and $30 < C < 300$. See Supplementary Information for a more detailed description of the theory. For $n^* < n < \approx 5 \times 10^{12}$/cm$^2$, we fit $\sigma(n)$ to Eqn. (1a) (figure 2a-dashed lines), to obtain the field-effect mobility $\mu_{FE} = Ce/n_{imp}h$ for each device. $\mu_{FE}$ ranges from 320 cm$^2$/Vs to 1500 cm$^2$/Vs reflecting different amounts of disorder in the samples. We identify the initial $n$-type dopants, and defects induced by mechanical exfoliation as likely sources of the disorder. The decrease in $\mu_{FE}$ with further electrolytic gating (device 1 run 2) indicates electrochemical damage, likely solvation of Se ions. The observation of sublinear $\sigma(n)$ at $n < n_{bulk}$ in devices 3-5 may indicate that there are additional types of disorder, e.g. neutral point defects, which need to be considered.

The observed minimum conductivity of the Dirac electronic band can be well understood through Eqn. (1b) as due to the residual carrier density $n^*$ in electron and hole puddles induced by the charged impurity potential at nominally zero carrier density: $\sigma_{min} = n^*e\mu$, where $n^*$ is calculated self-consistently[20] as a function of $n_{imp}$, $r_s$, and $d$, the distance of the impurities to the Dirac surface. The self-consistent theory predicts that $n^*$ increases with increasing disorder, and $\sigma_{min}$ depends only weakly on disorder. Figure 4 shows the experimentally observed residual carrier density $n^*$ per surface for each device (figure 4a) as well as $\sigma_{min}$ per surface (figure 4b) as a function of the experimentally-measured inverse mobility $1/\mu_{FE}$ which reflects the disorder strength. (For devices in which $n^*$ could be measured for $p$ and $n$-type conduction, both values are shown.) The shaded regions reflect the expectations of the self-consistent theory using parameter ranges $0.05 < r_s < 0.2$, and $d = 0.1$ Å to 15 Å. We see a good agreement between experiment and theory in that (1) $\sigma_{min}$ is weakly dependent on disorder strength ($1/\mu_{FE}$) and (2) $n^*$ increases with disorder strength ($1/\mu_{FE}$). Particularly for increasing disorder in the same device (Device 1 run 1 vs. run 2), $n^*$ increases but $\sigma_{min}$ is nearly unchanged (arrows in Figs. 4a and 4b). The experimental data agree best with the upper range of the theoretical estimates, corresponding to small $d = 0.1$ Å and large $r_s = 0.2$. Assuming $r_s = 0.2$, we infer an impurity density $n_{imp}$ ranging from 0.5 x $10^{13}$ cm$^{-2}$ to 2.3 x $10^{13}$ cm$^{-2}$, much larger than for graphene exfoliated on similar SiO$_2$ substrates [29], but comparable to the observed initial doping level of 1 x $10^{13}$ cm$^{-2}$ to 3 x $10^{13}$ cm$^{-2}$, suggesting that the dopants are the charged impurities responsible for limiting the mobility (see Supplementary Information).

The simple theory somewhat underestimates $n^*$ and $\sigma_{min}$, but we expect that the theory can be refined to take into account the non-linearity and asymmetry of the Bi$_2$Se$_3$ surface state bands [30]. Notably, the larger Fermi velocity for the electron band would increase the conductivity above the estimate in Eq. 1 for $n$-type conduction, indicating that disorder strength is likely somewhat underestimated from the $n$-type mobility. Shifting the points to the right (to larger disorder strength) in Fig. 4 would indeed improve the agreement between experiment and theory.

Reducing $n$-type doping of TI thin films by external agents provides an effective and simple way to probe topological surface transport properties in the absence of bulk conduction. For present devices the level of charged impurity disorder is in the order of ~$10^{13}$ cm$^{-2}$ limiting the mobility to 320 cm$^2$/Vs to 1500 cm$^2$/Vs. However, due to large dielectric constants in existing topological insulators, reduction of impurity concentrations to levels seen in the best bulk crystals (<$10^{17}$ cm$^{-3}$ corresponding to <$10^{11}$ cm$^{-2}$ in a 10 nm thick crystal[22]) would allow mobilities exceeding $10^5$ cm$^2$/Vs. Hence understanding and eliminating the doping presently observed in all thin crystals and films is of central importance to increasing the mobility of the topological surface state.

**Note added:** After submission of this work we became aware of a scanning tunneling microscopy study[10] which directly observed the screened potential fluctuations caused by charged impurity disorder in $Bi_2Se_3$ and $Bi_2Te_3$, consistent with our interpretation of the minimum conductivity arising from charge inhomogeneity.

**Methods**

$Bi_2Se_3$ thin crystals were produced by micro-mechanical cleavage of bulk $Bi_2Se_3$ single crystals and deposited on doped Si covered with 300nm $SiO_2$. Thin crystals with thickness about 10 nm were identified using atomic force microscopy (AFM). Thin film was patterned in Hall bar geometry using Ar plasma at a pressure of ≈6.7 Pa (5 x $10^{-2}$ Torr). Au/Cr electrodes were defined by electron-beam lithography (see inset of Fig. 1d). A brief (≈10 s) selective surface treatment of the contact area with $N_2$ or Ar plasma before the deposition of metals was used to enhance Ohmic conduction of the contacts.

*P*-type doping for devices 1-3 was achieved by applying negative voltage to a polymer electrolyte consisting of $LiClO_4$ and polyethylene oxide (PEO) in the weight ratio 0.12:1, as previously used for carbon nanotubes and graphene devices[24-26]. Molecular charge transfer doping for device 4 and 5 was done by thermal evaporation of ≈15 nm of F4TCNQ molecules (Aldrich) on top of the samples[27]. The devices were subsequently cooled down and further tuning of carrier density was done by sweeping the back gate voltage at cryogenic temperature. For electrolyte-gated measurements, the samples were cooled down to 250K in less than 1 min after applying the top gate voltage to minimize electrochemical reactions[25].

In addition to singly gated samples, we fabricated dual-gated samples based on F4TCNQ doped samples (Figure 2b). 60nm of hydrogen silsequioxane (HSQ, XR-1541, Dow Corning) was spin coated on a F4TCNQ coated $Bi_2Se_3$ device and top gate electrode was defined by electron beam lithography. We found that further fabrication on pre-doped devices increased the *n*-type doping level (e.g., ≈1.2 x $10^{13}$ $cm^{-2}$ at zero gate field in figure 2b compared to ≈0.3 x $10^{13}$ $cm^{-2}$ in figure 1d). From Hall carrier density vs. gate voltage, bottom and top gate capacitance were determined to be ≈11 $nF/cm^2$ and ≈33 $nF/cm^2$, respectively, which are reasonable considering the dielectric constants of $SiO_2$ ($\kappa \approx 3.9$) and HSQ ($\kappa \approx 3$).

Four-probe measurements of longitudinal and transverse electrical resistances were conducted using Stanford Research Systems SR830 Lock in amplifiers and a commercial cryostat equipped with 9 T superconducting magnet. Hall voltage was recorded in both polarities of the magnetic field (±1 T) and anti-symmetrized to remove longitudinal voltage components. In transport experiments a small and reproducible hysteresis in $V_{g,0}$ (≈2 V) was observed during forward and backward gate voltage scans. In consequence, resistivity and Hall data with same $V_g$ scan directions were compared in this work. Best fits to Eqn. (1a) were determined using a least squares linear fit to $\sigma(n)$ in the linear regime, determined by identifying the region of roughly constant slope $d\sigma/dn$. Thermal runs as described here were performed for more than ten different $Bi_2Se_3$ samples of similar thickness with qualitatively consistent results.

Acknowledgements

The study of electronic transport in novel materials during electrochemical modification is supported as part of the Science of Precision Multifunctional Nanostructures for Electrical Energy Storage, an Energy Frontier Research Center funded by the U.S. DOE, Office of Science, Office of Basic Energy Sciences under Award Number DESC0001160. Additional support was provided by NSF DMR-1105224. Preparation of $Bi_2Se_3$ was supported by NSF MRSEC (DMR-0520471) and DARPA-MTO award (N66001-09-c-2067). NPB was partially supported by the Center for Nanophysics and Advanced Materials. The authors acknowledge useful conversations with S. Das Sarma, E. Hwang, and D. Culcer.

Author Contributions

D. Kim conceived the p-type doping schemes. D. Kim and S. Cho fabricated devices, performed the electrical measurements with K. Kirshenbaum, and analyzed the data. N.P. Butch, P. Syers, and J. Paglione prepared single crystal $Bi_2Se_3$ starting material. S. Adam aided with the theoretical analysis. D. Kim, S. Cho, and M.S. Fuhrer wrote the manuscript.

Additional Information

The authors declare no competing financial interests. Supplementary information accompanies this paper. Correspondence and requests for materials should be addressed to Michael S. Fuhrer (mfuhrer@umd.edu).

Present Addresses

[†] Department of Physics and Materials Research Laboratory, University of Illinois at Urbana−Champaign, Urbana, Illinois 61801-2902, U.S.A.

‡ Condensed Matter and Materials Division, Lawrence Livermore National Laboratory, Livermore, CA 94550, U.S.A.

References

1. Fu, L., Kane, C. L. & Mele, E. J. Topological Insulators in Three Dimensions. *Phys. Rev. Lett.* **98**, 106803 (2007).
2. Zhang, H. J. *et al.* Topological insulators in $Bi_2Se_3$, $Bi_2Te_3$ and $Sb_2Te_3$ with a single Dirac cone on the surface. *Nature Phys.* **5**, 438-442 (2009).
3. Hsieh, D. *et al.* A tunable topological insulator in the spin helical Dirac transport regime. *Nature* **460**, 1101-1105 (2009).
4. Chen, Y. L. *et al.* Experimental Realization of a Three-Dimensional Topological Insulator $Bi_2Te_3$. *Science* **325**, 178-181 (2009).
5. Xia, Y. *et al.* Observation of a large-gap topological-insulator class with a single Dirac cone on the surface. *Nature Phys.* **5**, 398-402 (2009).
6. Zhang, T. *et al.* Experimental Demonstration of Topological Surface States Protected by Time-Reversal Symmetry. *Phys. Rev. Lett.* **103**, 266803 (2009).
7. Alpichshev, Z. *et al.* STM Imaging of Electronic Waves on the Surface of $Bi_2Te_3$: Topologically Protected Surface States and Hexagonal Warping Effects. *Phys. Rev. Lett.* **104**, 016401 (2010).
8. Hanaguri, T. *et al.* Momentum-resolved Landau-level spectroscopy of Dirac surface state in $Bi_2Se_3$. *Phys. Rev. B* **82**, 081305R (2010)
9. Cheng, P. *et al.* Landau Quantization of Topological Surface States in $Bi_2Se_3$. *Phys. Rev. Lett.* **105**, 076801 (2010)
10. Beidenkopf, H. *et al.* Spatial fluctuations of helical Dirac fermions on the surface of topological insulators. *Nature Phys.* **7**, 939–943 (2011)
11. Steinberg, H., Gardner, D. R., Lee, Y. S. & Jarillo-Herrero, P. Surface State Transport and Ambipolar Electric Field Effect in $Bi_2Se_3$ Nanodevices. *Nano Lett.* **10**, 5032-5036 (2010).
12. Qu, D., Hor, Y. S., Xiong, J., Cava, R. J., & Ong, N. P. Quantum Oscillations and Hall Anomaly of Surface States in the Topological Insulator $Bi_2Te_3$. *Science* **329** 821-824 (2010)
13. Xiu, F. et al. Manipulating Surface States in Topological Insulator Nanoribbons. *Nature Nanotech.* **6**, 216-221 (2011)
14. Analytis, J. G. *et al.* Two-dimensional surface state in the quantum limit of a topological insulator. *Nature Phys.* **6**, 960-964 (2010).
15. Peng, H. *et al.* Aharonov-Bohm interference in topological insulator nanoribbons. *Nature Mater*. **9**, 225-229 (2010).
16. Chen, J. *et al.* Gate-Voltage Control of Chemical Potential and Weak Antilocalization in $Bi_2Se_3$. *Phys. Rev. Lett*. **105**, 176602 (2010).
17. Checkelsky, J.G. Hor, Y.S., Cava R. J. & Ong N. P. Surface state conduction observed in voltage-tuned crystals of the topological insulator $Bi_2Se_3$. *Phys. Rev. Lett*. **106**, 196801 (2010).
18. Novoselov, K. S. *et al.* Two-dimensional gas of massless Dirac fermions in graphene. *Nature* **438**, 197-200 (2005).
19. Hwang, E. H., Adam, S. & Das Sarma, S. Carrier Transport in Two-Dimensional Graphene Layers. *Phys. Rev. Lett.* **98**, 186806 (2007).
20. Adam, S., Hwang, E. H., Galitski, V. M. & Das Sarma, S. A self-consistent theory for graphene transport. *P. Natl. Acad. Sci. USA* **104**, 18392-18397 (2007).
21. Culcer, D., Hwang, E. H., Stanescu, T. D. & Das Sarma, S. Two-dimensional surface charge transport in

topological insulators. *Phys. Rev. B* **82**, 155457 (2010).
22. Butch, N. P. *et al.* Strong surface scattering in ultrahigh-mobility $Bi_2Se_3$ topological insulator crystals. *Phys. Rev. B* **81**, 241301 (2010).
23. Kong, D. *et al.* Rapid Surface Oxidation as a Source of Surface Degradation Factor for $Bi_2Se_3$. *ACS Nano* **5**, 4698-4703 (2011).
24. Das, A. *et al.* Monitoring dopants by Raman scattering in an electrochemically top-gated graphene transistor. *Nature Nanotech.* **3**, 210-215 (2008).
25. Efetov, D. K. & Kim, P. Controlling Electron-Phonon Interactions in Graphene at Ultrahigh Carrier Densities. *Phys. Rev. Lett.* **105**, 256805 (2010).
26. Lu, C. G., Fu, Q., Huang, S. M. & Liu, J. Polymer electrolyte-gated nanotube field-effect carbon transistor. *Nano Lett.* **4**, 623-627 (2004).
27. Coletti, C. *et al* Charge neutrality and band-gap tuning of epitaxial graphene on SiC by molecular doping. *Phys. Rev. B* **81**, 235401 (2010).
28. Bansal, M., Kim Y. S., Brahlek, Eliav Edrey, E., & Oh, S. Thickness-independent surface transport channel in topological insulator $Bi_2Se_3$ thin films. *arXiv:1104.5709* (2011).
29. Chen, J. H. *et al* Charged Impurity Scattering in Graphene. *Nature Phys.* **4**, 377-381 (2008).
30. Adam, S., Hwang, E. H. & Sarma, S. D. 2D transport and screening in topological insulator surface states. *arXiv:1201.4413* (2012).

**Figure captions**

**Figure 1 $Bi_2Se_3$ thin film device.** Schematics of *p*-type doping scheme and gate configuration for **a,** charge transfer doping with F4TCNQ organic molecules and **b,** polymer electrolyte ($PEO+LiClO_4$) top gating. **c**, Longitudinal resistivity $\rho_{xx}$ and **d**, sheet carrier density determined from Hall measurement as a function of back gate voltage for device 4 (F4TCNQ doped) at various temperatures from 2 K to 50 K indicated in caption. The inset shows an optical micrograph of the device. The scale bar is 2μm.

**Figure 2 Single band conduction in the topological insulator regime. a,** Hall resistivity $\rho_{xy}$ of device 4 as a function of magnetic field *B* at a temperature of 2 K at different carrier densities tuned by the back gate electrode. **b,** Polar plot of the normalized longitudinal resistivity $\rho_{xx}$ of dual gated $Bi_2Se_3$ thin film device as a function of total magnitude of displacement field ($D_{total}$) and gating asymmetry factor α (defined in the text)

**Figure 3 Transport properties of $Bi_2Se_3$ surface state. a**, The conductivity per surface vs. carrier density per surface σ(*n*) at zero magnetic field for five different devices. Device 1 to 3 are electrolyte gated and device 4 and 5 are F4TCNQ doped. The inset shows σ(*n*) near the Dirac point. Dashed lines are fits to Eqn. (1a). Transport data outside the topological regime ($n > n_{bulk} = 5 \times 10^{12}/cm^2$) are denoted as dotted curves. **b**, Hall carrier density per surface vs. carrier density measured at the same conditions as in (a). Dashed lines show residual carrier density *n*\* (defined in the text) for different devices **c**, Variation of field effect mobility as a function of carrier density. Dashed curves indicate the region |*n*| < *n*\* within which electron and hole puddles dominate transport.

**Figure 4 Minimum conductivity and charge inhomogeneity vs. inverse mobility. a**, Residual carrier density *n*\* vs. inverse field effect mobility (1/$μ_{FE}$) and **b**, minimum conductivity $σ_{min}$ vs. 1/$μ_{FE}$. Shaded areas indicate the expectations of the self-consistent theory of Ref. 20, open symbols are experimental data.

Figure 1

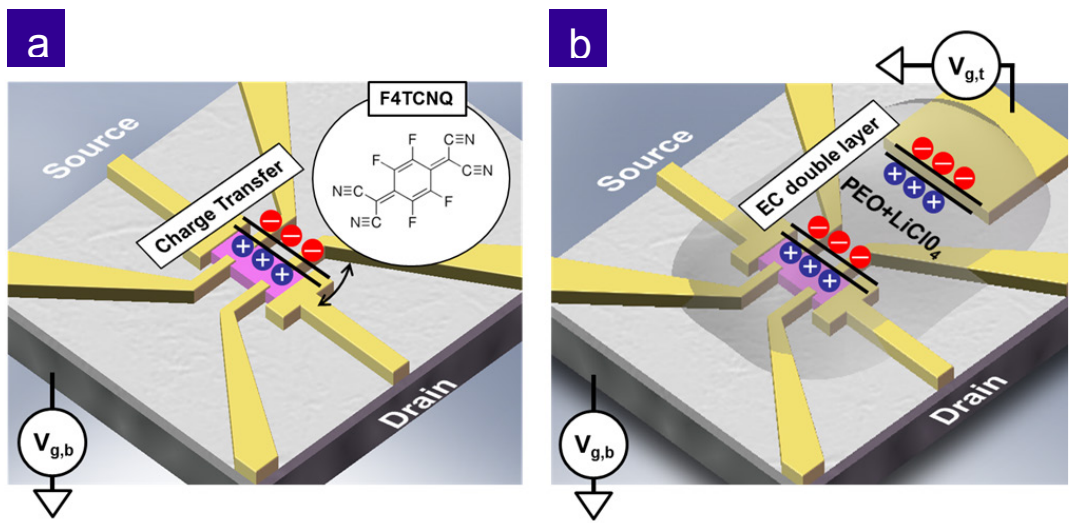

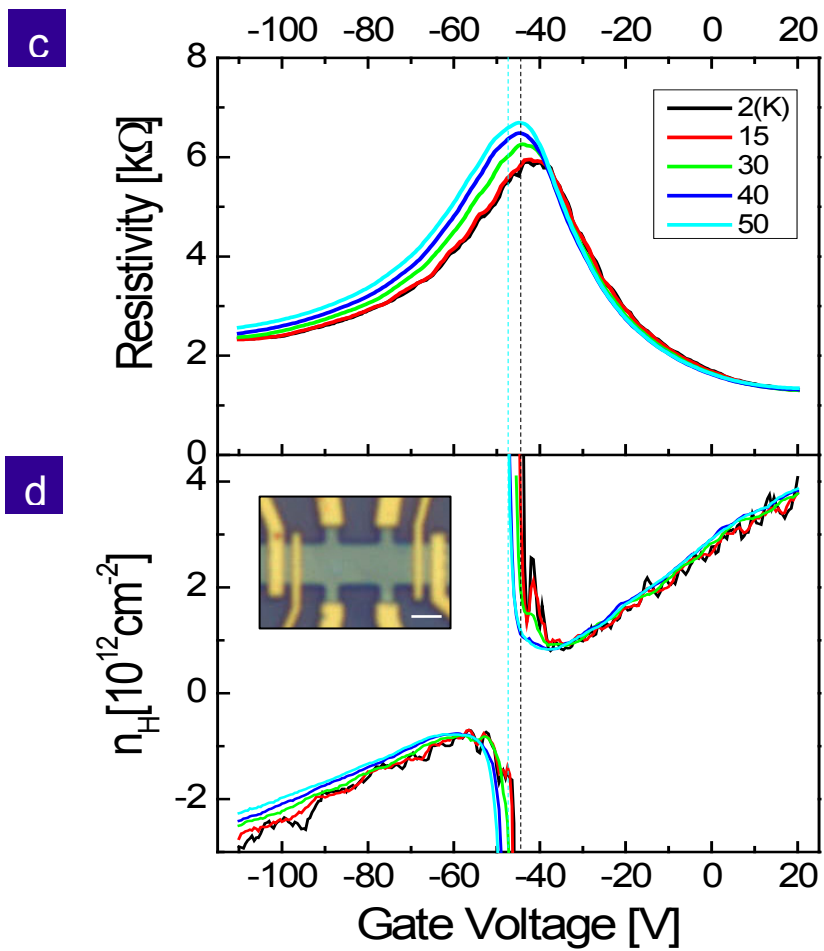

Figure 2

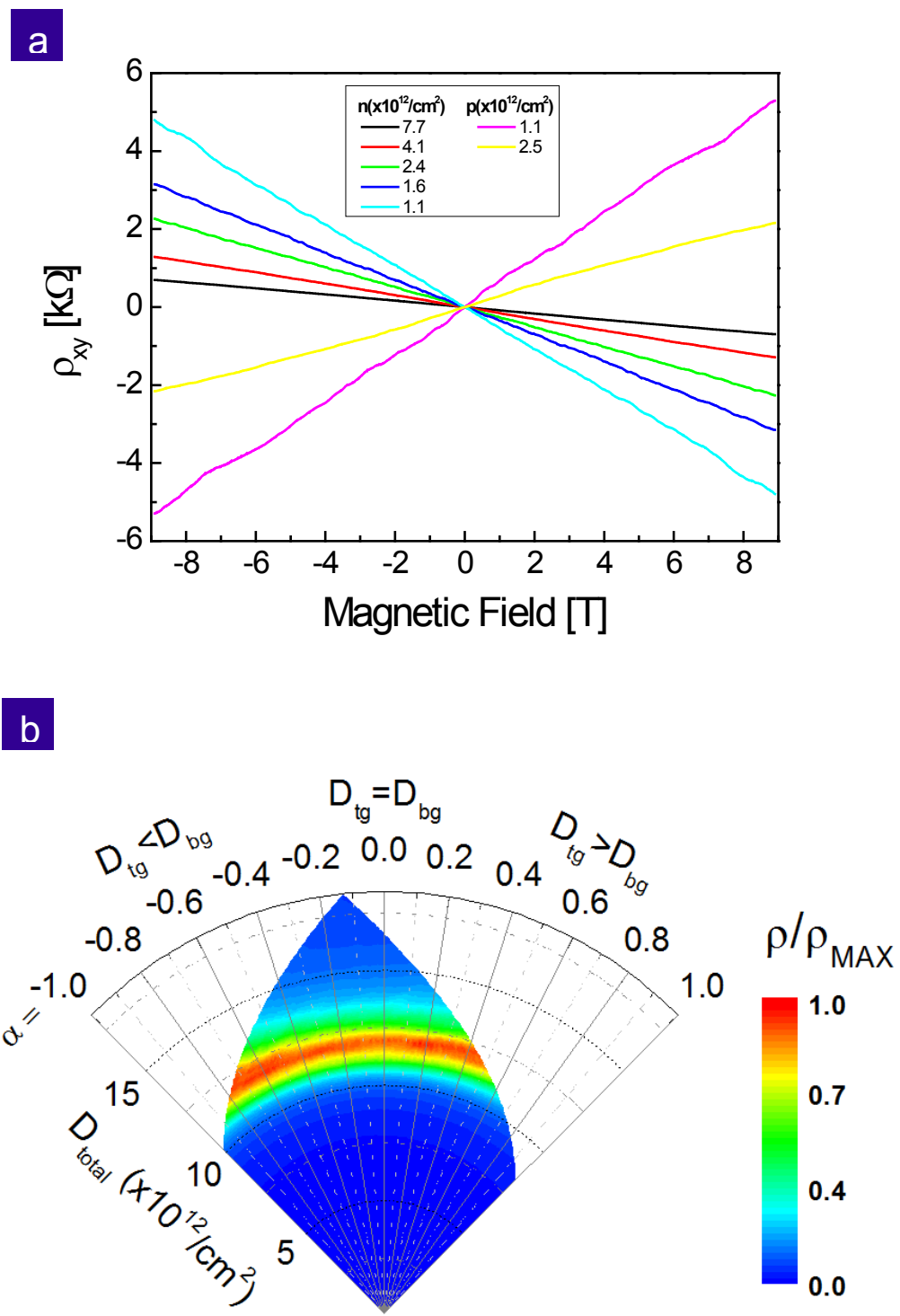

Figure 3

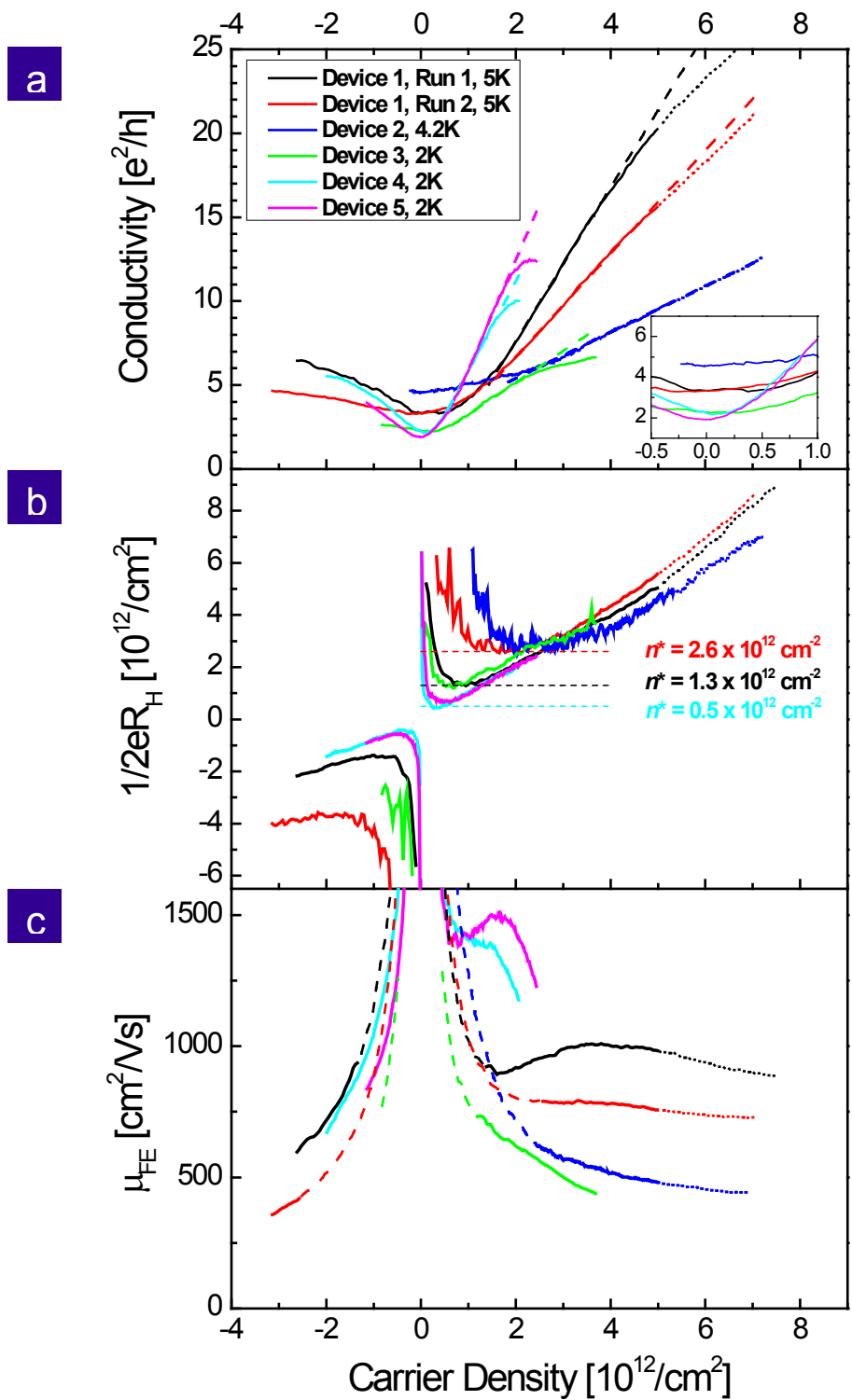

Figure 4

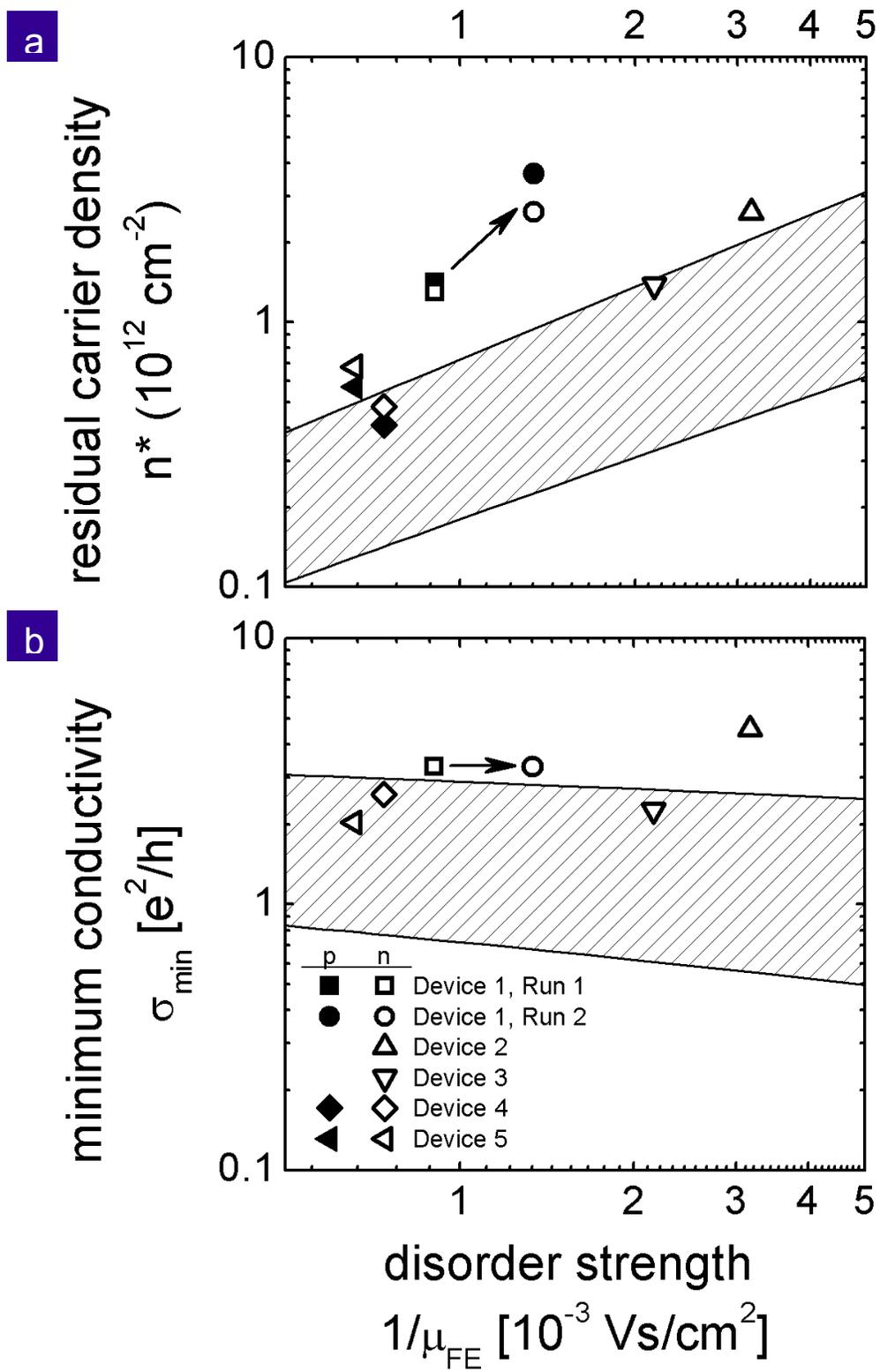

Surface conduction of topological Dirac electrons in bulk insulating Bi$_2$Se$_3$

Dohun Kim*[1], Sungjae Cho*[1], Nicholas P. Butch[1], Paul Syers[1], Kevin Kirshenbaum[1], Shaffique Adam[2], Johnpierre Paglione[1], and Michael S. Fuhrer[1]+

1. Center for Nanophysics and Advanced Materials, Department of Physics, University of Maryland, College Park, MD 20742-4111, USA

2. Center for Nanoscale Science and Technology, National Institute of Standards and Technology, Gaithersburg, MD 20899-6202, USA

**Supplementary Information**

**1. Interlayer electrostatic coupling and simultaneous tuning of bottom and top surfaces**

We discuss possible origin of observed simultaneous gating of topological surfaces (figure 2b in the main text). We model the electrostatic response of the two topological surfaces to back and top gate voltage by the effective capacitor network as shown in figure S1a, which includes the electrostatic gate capacitances $C_{g,t}$ and $C_{g,b}$ to top and bottom surfaces respectively, and the quantum capacitances $C_{q,t}$ and $C_{q,b}$ of top and bottom surfaces. Note that the two surfaces are coupled by an interlayer capacitance $C_{il}$, which originates from the bulk part of the Bi$_2$Se$_3$ thin film. For the dual-gated device 5, we measure $C_{g,b}$ = 11 nF/cm$^2$, $C_{g,t}$ = 33 nF/cm$^2$, and estimate $C_{il}$ to be order of ~10 uF/cm$^2$ considering nominal thickness (≈10 nm) and dielectric constant ($\kappa \approx$ 100 [1]) of Bi$_2$Se$_3$ films. For electrolyte-gated devices, we measure a capacitance to the back gate at cryogenic temperatures which is greater than 11 nF/cm$^2$; we assume that the excess capacitance is due to a high polarizability of the polymer electrolyte which provides additional coupling of the back gate to the top surface on order 16 nF/cm$^2$. The near-equivalence of the capacitance of the back gate to top and bottom surfaces in electrolyte-gated devices further enhances equilibration of charge on top and bottom surfaces above what would be expected in the analysis below.

The quantum capacitance reflects the change in charge required to change the Fermi energy $E_F$ in a material. For small changes in Fermi energy, $C_q$ Is often approximated by $e^2(dn/dE_F)$, where $e$ is the elementary charge and $dn/dE_F$ is the density of electronic states, Since $D(E)$ varies significantly in a 2D Dirac material such as the topological surface state, we will use the full quantum capacitance $C_q = e^2(n/E_F)$ where $n$ is the carrier density. We assume a linear dispersion relation for surface states in Bi$_2$Se$_3$ which is valid up to $n \approx 5 \times 10^{13}$/cm$^2$ per surface[2], thus considering degeneracy of one, $E_{Fb(t)} = 2\hbar v_F(\pi n_{b(t)})^{1/2}$ where $\hbar$ is Planck's constant and $v_F = 3 \times 10^5$ m/s to $7 \times 10^5$ m/s is the Fermi velocity in Bi$_2$Se$_3$, representing roughly the average Fermi velocity for the conduction cone for energies below the bulk conduction band[3-4]. We use the experimentally observed $n_0 \approx 12 \times 10^{12}$/cm$^2$ at zero displacement field as a reference point; equal carrier density ($n_t = n_b = 6 \times 10^{12}$/cm$^2$) at this gate voltage was assumed. The carrier density of each surface can be expressed as

$$n_{b(t)} = n_{bg(tg)} \mp n_{il} \qquad (1)$$

where $n_{bg(tg)} = C_{g,b(t)}(eV_g - E_{Fb(t)})/e^2$, $n_{il} = C_{il}(E_{ft} - E_{fb})/e^2$ and $V_g$ is the applied back gate voltage.

Figure S1b shows calculated carrier density for bottom and top surface carrier densities as functions of total magnitude of applied field for symmetric ($\alpha = 0$, dashed, asymmetry factor $\alpha$ is defined in the main text) and highly asymmetric ($\alpha = -0.5$) gating schemes assuming $C_{g,b} = 11$ nF/cm$^2$ and $C_{g,t} = 33$ nF/cm$^2$ corresponding to device 5. We compare carrier densities for the same parameters for gate capacitances in the strong ($C_{il}/C_q \gg 1$, $C_{il} = 30$ uF/cm$^2$, solid lines), intermediate ($C_{il}/C_q \sim 1$, $C_{il} = 3$ uF/cm$^2$, dotted lines), and weak coupling limit ($C_{il}/C_q \ll 1$, $C_{il} = 0.1$ uF/cm$^2$, dashed-dot lines). In the weak coupling limit the large difference in applied field ($D_{gb} = 3D_{gt}$ for $\alpha = -0.5$) leads to very different carrier densities on either side of the device so that a splitting of maximum resistivity at Dirac point is expected. However, the presence of large inter-surface capacitance enhances the tendency to equalize the surface charge densities. As interlayer capacitance becomes comparable to quantum capacitance of Dirac surface, it dramatically reduces carrier density difference of the surfaces.

For estimated $C_{il} \approx 11$ µF/cm$^2$ of device 5 ($t_{tot} \approx 9$ nm), the carrier density difference at charge neutrality $\Delta n = n_t - n_b \approx 1.2 \times 10^{12}$ cm$^{-2}$, which is comparable to the observed width of charge inhomogeneity ($2n^*$). However, we note that the surface electronic state has some finite extent into the bulk, thus especially in thin film geometry, $C_{il}$ can be significantly larger. For $C_{il} > 13$ µF/cm$^2$, $\Delta n$ becomes less than $2n^*$ ($\approx 1 \times 10^{12}$ cm$^{-2}$), and for $C_{il}$ exceeding 20 µF/cm$^2$ which is conceivable when $t_{bulk} < 5$ nm, Fermi energies of bottom and top surfaces are essentially locked together with negligible $\Delta n$. The experimental observations in the main text suggest we approach this limit; we observe that the simultaneous gating approximation holds even when $\alpha = \pm 1$, where gating field is applied only to one of the surfaces. Since $\Delta n$ is much less than the observed $n^*$ in this limit, we conclude that $n^*$ is dominated by local inhomogeneity in the carrier density due to disorder. Therefore, we can parameterize the transport properties for singly gated devices (figure 3 in the main text) as functions of carrier density *per* surface $n$ estimated from $n = (C_g/2e)(V_g - V_{g,0})$ where $C_g$ is estimated from the slope of $n_H(V_g)$ in the unipolar regime.

## 2. Weak anti-localization of $Bi_2Se_3$ surface states.

We discuss the behavior of low field magnetotransport in the TI regime. Figure S2 shows the magneto-conductivity per layer $\Delta\sigma(H)$ of device 1 (run 1) at various carrier densities. $\Delta\sigma(H)$ curves at high $n > 5 \times 10^{12} cm^{-2}$ are nearly identical, but change significantly as the Fermi energy approaches the Dirac point[5,6]. In the limit of strong spin orbit interaction and low mobility, we can fit the low field magneto-conductance per surface to the simplified form of the theory for weak anti-localization[7],

$$\Delta\sigma(H) = A\frac{e^2}{h}[\ln\frac{H_0}{H} - \psi\left(\frac{1}{2} + \frac{H_0}{H}\right)] \quad (2)$$

where $H_0 = \hbar/4De\tau_0$ is a characteristic field related to the dephasing time $\tau_0$, $\psi$ is the digamma function, and A is a fit parameter whose theoretically expected value is $1/2\pi$. Confining the fit to fields below 0.4T, Eqn. (2) does provide reasonably good fits to the data at all carrier densities. Fitting results across the carrier density range are shown in the inset of figure S2. At the charge neutrality point the coefficient A rises to $\approx 0.2$ in the topological regime, which is within 25% of the theoretical value $A = 1/2\pi$ [7]. The observation of magnetoconductivity described by Eqn. (2) in the clear absence of bulk carriers confirms its surface state origin, as posited in Refs. 5 and 6. $H_0$ is maximum (indicating the coherence length is minimum) near $n = 0$; similar behavior is observed in the electron-hole puddle regime in graphene[8].

## 3. Conductivity of STIs limited by charged impurity scattering

Charged impurities can be present in the $Bi_2Se_3$ bulk, in the polymer electrolyte, or physisorbed onto the surface. For Dirac fermions, such long-range Coulomb disorder is expected to give signatures in the transport measurements at both high carrier density and at low carrier density. Away from the Dirac point, the conductivity limited by charged impurity scattering is linear in the carrier density i.e. $\sigma(n) = ne\mu$ where the mobility $\mu$ is calculated from the Boltzmann transport theory[2,9]. Close to the Dirac point, the same charged impurities locally shift the Dirac point creating electron and hole puddles. The non-zero carrier density in the puddles gives rise to a minimum conductivity $\sigma_{min} = n^* e \mu$, where $n^*$ is the characteristic density of carriers inside the puddles. In our experiments on $Bi_2Se_3$ we observe both these signatures of charged impurities. More importantly, we can measure $\sigma_{min}$, $n^*$ and $\mu$ as independent experimental parameters, and test the charged impurity picture by checking to see if the variation of $\sigma_{min}$ and $n^*$ with $\mu$ is consistent with theoretical expectations.

The self-consistent theory has three parameters: the density of charged impurities $n_{imp}$, the effective interaction parameter (dimensionless Wigner-Seitz radius) $r_s = e^2/\kappa_{eff}\hbar v_f$, where $\kappa_{eff}$ is average dielectric constant, and the distance $d$ of the impurities from the surface. For simplicity we ignore the weak dependence of $\mu$ on $d$, and assume that the surface bands have constant $v_F$. In this case, we find

$$\frac{\mu}{\mu_0} = \frac{5}{4r_s^2} \frac{n_0}{n_{imp}} \left[ \frac{\pi}{4} + \frac{3r_s}{2} - \frac{3r_s^2\pi}{8} + \left( \frac{3r_s^3}{4} - 2r_s \right) \frac{\arccos(2/r_s)}{\sqrt{r_s^2 - 4}} \right]^{-1} \quad (3)$$

where $\mu_0 = 1 \, m^2/Vs$ and $n_0 = 10^{10} cm^{-2}$. Following Ref. 9, $n^*$ is calculated self-consistently. Using the Thomas-Fermi screening theory, we have

$$\frac{n^*}{n_{imp}} = \frac{r_s^2}{2} \left[ -1 + e^{4r_s d\sqrt{\pi n^*}} \left( 1 + 4r_s d\sqrt{\pi n^*} \right) \Gamma_0 \left( 4r_s d\sqrt{\pi n^*} \right) \right] \quad (4)$$

where $\Gamma_0$ is the exponential integral function $\Gamma_0(x) = \int_x^\infty t^{-1} e^{-t} dt$. Given $n^*$ and $\mu$, we can calculate $\sigma_{min} = n^* e \mu$. Due to uncertainty in $v_f$ and $\kappa$, in Fig. 4 of the main text, we show the window of theoretical results obtained for $0.1$ Å $< d < 15$ Å and $0.05 < r_s < 0.2$. The $r_s$ range corresponds to the existing estimates in the literature of $3 < v_f < 7 \times 10^5$ m/s [3,4], and $30 < \kappa_{eff} < 60$ [1,10].

Note that at high density, when $n >> n^*$, the result in Eq. 3 differs by a factor of ≈4 from that of Ref. [2]. At low density, our finding that $n^* << n_{imp}$ (Eq. 4) is an important prediction of the self-consistent theory. If $n^*$ was of the same order as $n_{imp}$ as assumed in Ref. [2] then the disorder induced energy fluctuations would be comparable to the band-gap and we would have been unable to observe the minimum conductivity and other surface band properties.

### 4. Gate tuned transport of Bi$_2$Se$_3$ device 1 with polymer electrolyte at 300K

Figure S3 shows resistivity (ρ) and Hall carrier density ($n_H$) as a function of electrolyte gate voltage measured at 300K. Although the initial gate capacitance for the polymer electrolyte was found to be ≈1 µF/cm$^2$, the gate efficiency decreased rapidly as the gate voltage exceeded -0.8V. Notable hysteresis in the forward and backward voltage scans was also found. However, we note that the possible mechanism which can cause hysteresis (e.g. polymer decomposition or reaction with the sample) is reversible in this voltage scan range (0V to 1V). Because of additional doping induced by mechanical cleavage and reaction with ambient species[11,12], the amount of total carrier density at zero gate voltage was found to be much greater than the expected considering the bulk charge density (≈10$^{17}$ cm$^{-3}$) in our low-doped starting material[1]. The observed initial doping level (≈3.1x 10$^{13}$ cm$^{-2}$) is of similar magnitude to the $n_{imp}$ obtained by fit to the theory of charged impurity governed conductivity in $Bi_2Se_3$[2] indicating that the dopants act as charged impurities.

## 5. Alumina ($Al_2O_3$) capped $Bi_2Se_3$ device

Motivated by recent reports[11,12] showing surface degradation of $Bi_2Se_3$ upon exposure to atmosphere, we performed experiments of capping the surface with $Al_2O_3$. Thin (10nm) $Bi_2Se_3$ was mechanically exfoliated onto highly n-doped $SiO_2$ (300nm)/Si in a glove box of nitrogen environment, transferred to an e-beam evaporator without exposing to ambient environment. Subsequently, very thin (1.5nm) alumina ($Al_2O_3$) was deposited at a base pressure of $\sim10^{-5}$ Pa ($\sim10^{-7}$ Torr) and the thin capping layer prevented $Bi_2Se_3$ surface from being exposed to air afterwards. Electrodes (Cr/Au) were prepared and electrical measurement with dual gate was done in the same way as described in the main text. Figure S4 shows conductivity and Hall carrier density measured at low temperature (4.2K). While we found it was possible to gate the sample through the Dirac point, we found Hall mobility of $Al_2O_3$ capped device $\approx$100 $cm^2$/Vs (3 to 15 times lower than uncapped device of the same thickness), residual density $\approx$3 x $10^{13}$ $cm^{-2}$ (8 to 80 times higher, and higher than the density at which bulk bands are populated). Following the analysis in the main text, we would extract a charged impurity density $\approx3\times10^{14}$ $cm^{-2}$ (5 to 10 times higher), though the analysis is beyond its range of validity since the bulk bands are significantly populated. The result indicates that the $Al_2O_3$ capping layer itself may degrade the surface states by providing strong source of Coulomb charged impurities although capping surface may prevent further degradation of the surface from air exposure.

# References

1. Butch, N. P. *et al.* Strong surface scattering in ultrahigh-mobility $Bi_2Se_3$ topological insulator crystals. *Phys Rev B* **81**, 241301 (2010).
2. Culcer, D., Hwang, E. H., Stanescu, T. D. & Das Sarma, S. Two-dimensional surface charge transport in topological insulators. *Phys. Rev. B* **82**, 155457 (2010).
3. Kuroda, K. et al. Hexagonally Deformed Fermi Surface of the 3D Topological Insulator $Bi_2Se_3$. *Phys. Rev. Lett.* **105**, 076802 (2010).
4. Zhu, Z. H. et al. Rashba spin-splitting control at the surface of the topological insulator $Bi_2Se_3$. *arXiv:1106.0552* (2011).
5. Chen, J. et al. Gate-Voltage Control of Chemical Potential and Weak Antilocalization in $Bi_2Se_3$. *Phys. Rev. Lett.* **105**, 176602 (2010).
6. Checkelsky, J.G.  Hor, Y.S., Cava R. J. & Ong N. P. Surface state conduction observed in voltage-tuned crystals of the topological insulator $Bi_2Se_3$. *Phys. Rev. Lett.* **106**, 196801 (2010).
7. S. Hikami, A. I. L., Y. Nagaoka. Spin-Orbit Interaction and Magnetoresistance in the Two Dimensional Random System. *Prog. Theor. Phys.* **63**, 707-710 (1980).
8. Gorbachev, R. V., Tikhonenko, F. V., Mayorov, A. S., Horsell, D. W. & Savchenko, A. K. Weak localization in bilayer graphene. *Phys. Rev. Lett.* **98**, 176805 (2007).
9. Adam, S., Hwang, E. H., Galitski, V. M. & Das Sarma, S. A self-consistent theory for graphene transport. *P. Natl. Acad. Sci. USA* **104**, 18392-18397 (2007).
10. Beidenkopf, H. et al. Spatial fluctuations of helical Dirac fermions on the surface of topological insulators. *Nature Phys.* **7**, 939–943 (2011).
11. Kong, D. et al Rapid Surface Oxidation as a Source of Surface Degradation Factor for $Bi_2Se_3$. *ACS Nano* **5**, 4698-4703 (2011).
12. Analytis, J. G. et al. Two-dimensional surface state in the quantum limit of a topological insulator. *Nature Phys.* **6**, 960-964 (2010).

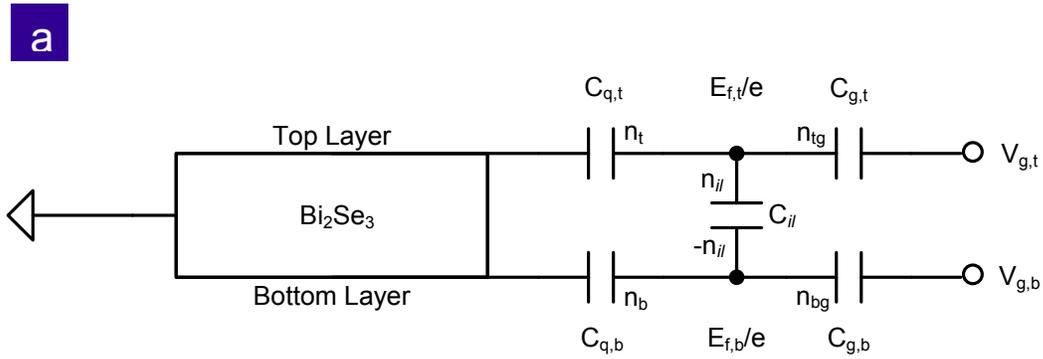

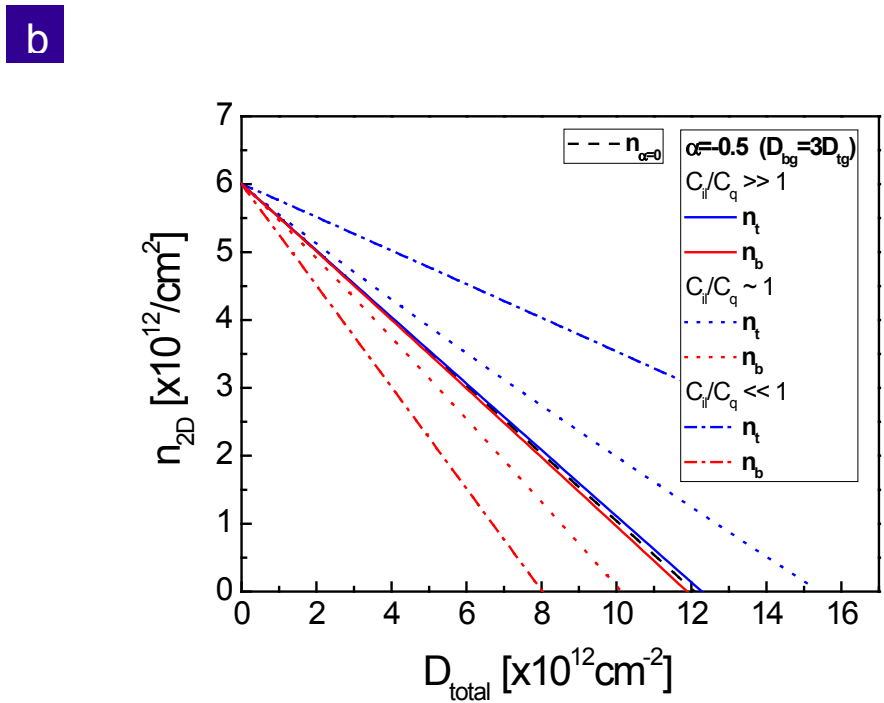

**Figure S1 Displacement field induced surface charge density. a,** Capacitor network model of Bi$_2$Se$_3$ thin film device. **b,** Calculated charge densities for top (blue) and bottom (red) surfaces as a function of total magnitude of applied field. Dashed black line is carrier density for symmetric applied field ($\alpha=0$). The solid, dot, dashed-dot lines are charge densities calculated for highly asymmetric applied field ($\alpha=-0.5$) in the strong, intermediate, and weak interlayer coupling limits, respectively.

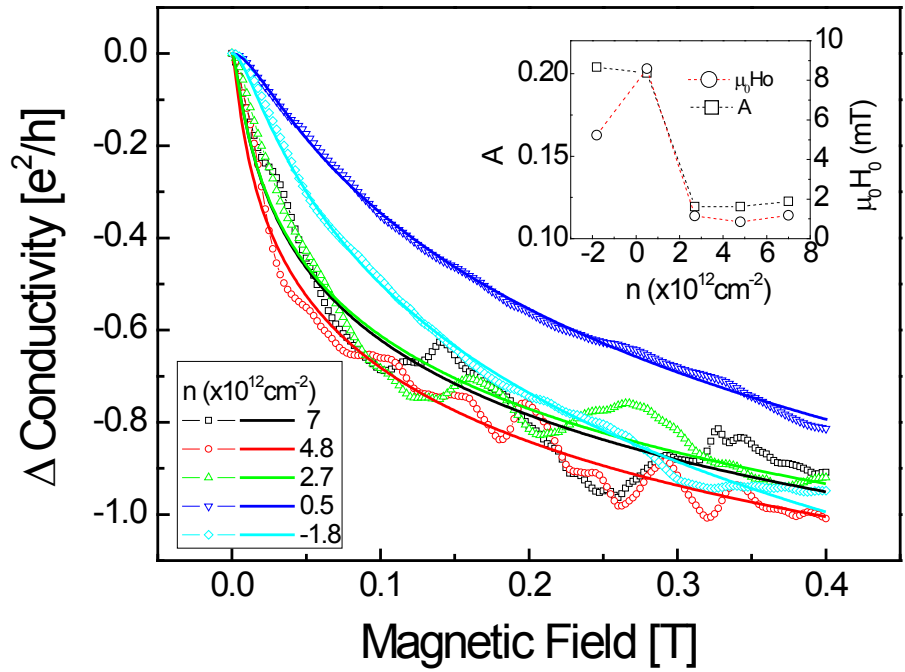

**Figure S2 Magnetic field suppression of weak anti-localization.** Symmetrized conductivity change of device 1 (symbols) vs. magnetic field at various per surface carrier densities. Solid curves are fit to Eqn. (2). The inset shows prefactor $A$ (squares) and dephasing field $H_0$ (circles) obtained from fits to Eqn. (2).

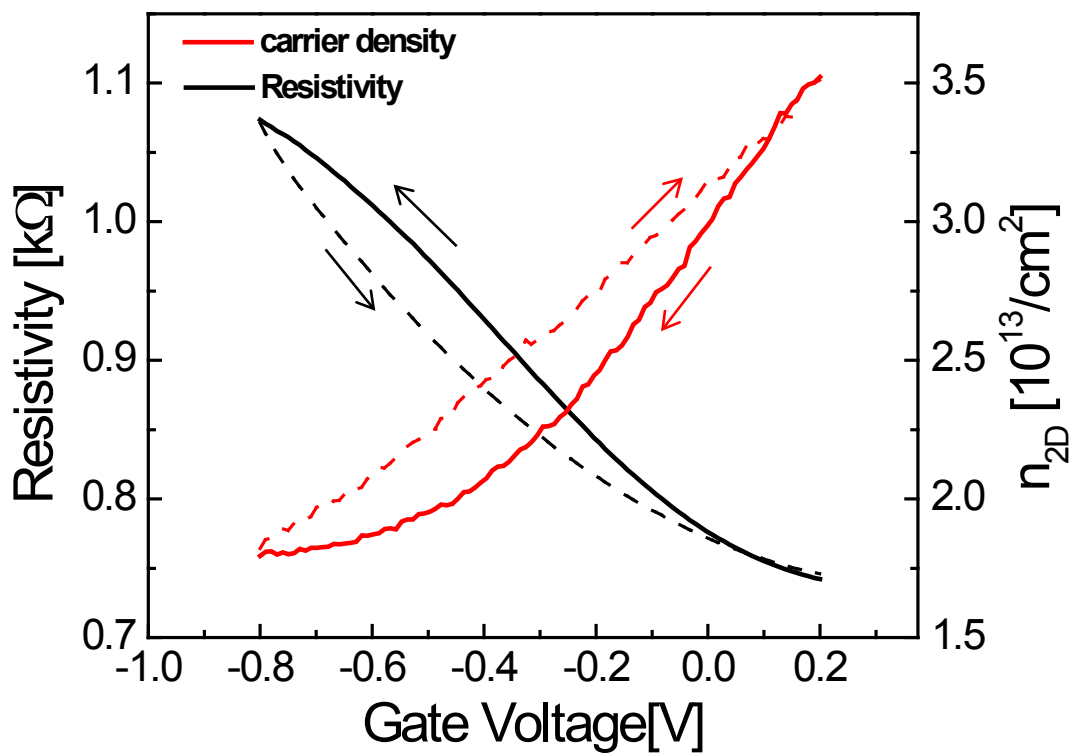

**Figure S3 Electrolyte gating of device 1 at 300K.** Resistivity (ρ) and sheet carrier density determined from Hall voltage ($n_H$) as a function of electrolyte gate voltage.

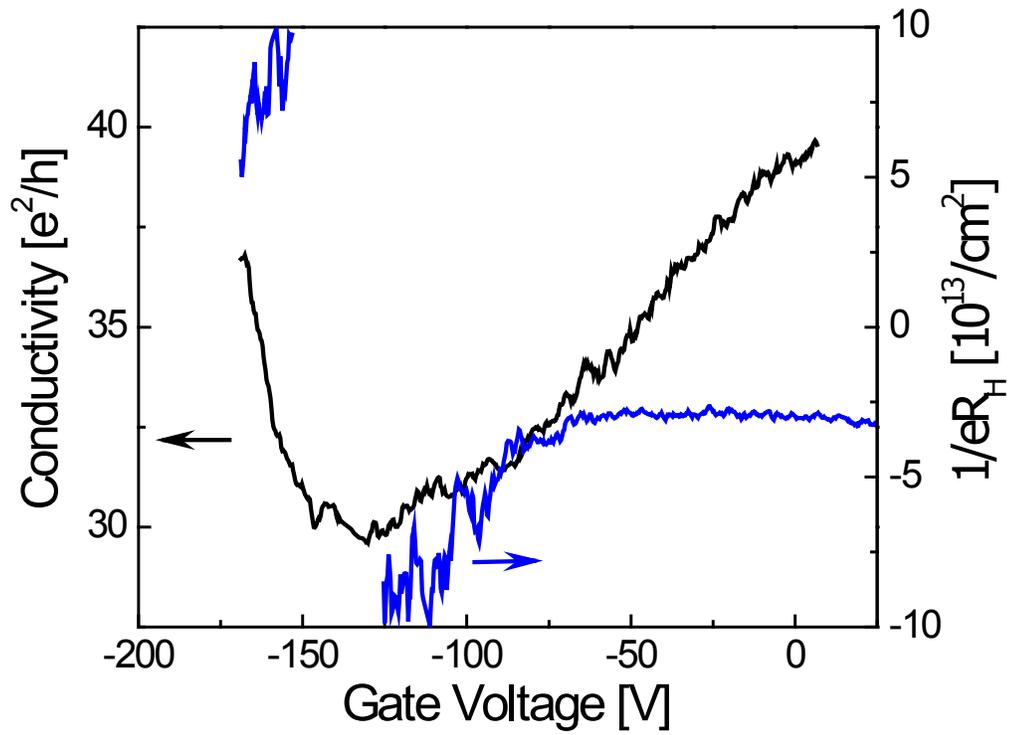

**Figure S4 Electronic transport of Al$_2$O$_3$ capped Bi$_2$Se$_3$ device.** Conductivity (σ) and Hall carrier density measured in an Al$_2$O$_3$ capped device at T=4.2K as a function of back gate voltage.